\documentclass[12pt]{article}
\usepackage{graphicx}
\usepackage{amsmath}
\usepackage{psfrag}
\usepackage{amssymb}
\usepackage{chicago}

\newcommand{\mcite}[1]{{\bf [\citeNP{#1}]}}

\begin{document}






\title{Random Walker Ranking for\\ NCAA Division I-A Football}

\author{Thomas Callaghan, Peter J. Mucha,
  and Mason A. Porter}
\date{}  \maketitle

\section{INTRODUCTION.}    

The Bowl Championship Series (BCS) agreement was created in 1998
to match the top two NCAA Division I-A college football teams
in a National Championship game at the end of each season. As part of the
agreement, the official BCS Standings are used to pick which two teams
most deserve to play in the championship game and which teams should
play in the other major bowl games. The BCS Standings have
significant financial ramifications, as the 
National Championship Game and major bowl games yield financial payouts
to the conferences of the appearing teams projected to be \$14--\$17 million per team in 2007 \mcite{BCS}.
In addition to this direct financial benefit, BCS bowl appearances
likely generate indirect gains from increases in both alumni contributions and student applications.  Before the BCS, the matchups in many bowl games were determined according to
conference traditions, so matches between the \#1 and \#2 teams in the
nation rarely occurred.  At times, this yielded multiple undefeated
teams and co-National Champions (most recently, Nebraska and Michigan
in 1997).  On other occasions, a single team with an arguably easy
schedule could go undefeated and be declared National Champion by
polls without ever having played a ``major'' opponent (e.g., BYU in
1984).

The BCS system endeavors to address these problems while maintaining
the tradition of finishing the season with bowl games. Prior to 2004,
BCS Standings were determined by a combination of two polls (coaches
and sportswriters), selected algorithmic rankings, strength of
schedule, losses, and bonus points awarded for defeating highly-ranked
teams. This system double-counted key contributions, because
reasonable polls and rankings already took losses and strength of
schedule into account (see, for example, \mcite{Callaghan04}).  In
2004, the BCS instituted a formula that simply averages polls and
computer rankings, but the system continues to vary from year to year as specific polls and computer systems are added and removed.

The fundamental difficulty in accurately ranking or even agreeing on a
ranking methodology for college football lies in two factors---the
dearth of games played by each team and the large disparities in types
and difficulties of individual schedules.  For instance, should
an undefeated team from a purportedly weaker conference with few tough
nonconference opponents be ranked ahead of a ``major conference''
team that lost a game or two while playing difficult opposition both
in its conference and in its nonconference matchups?  While each game 
outcome is an imperfect paired comparison between two teams,
the 10--13 regular season games
(including conference championships) played by each of the 119 
Division I-A football teams severely
limits the quantity of information relative to, for example, college
and professional basketball and baseball schedules.  Moreover, most of
the Division I-A football teams play the majority of their games
within their conferences, and there are significant variations in the level of
play across different conferences that further complicate attempts to select
the top two teams from the available information.  To make matters
worse, it is not even clear what the phrase ``top two'' should mean: does it
refer to the two teams with the best overall seasons or the two
playing the best at the end of the season?

Despite the obvious difficulties, many systems for ranking college
football teams have been promoted by mathematically and
computationally inclined fans
(see, for example, those listed by Massey \mcite{MasseyComparisons}).  
Many of these schemes are
relatively complicated mathematically, making it virtually impossible
for the lay sports enthusiast to understand the ranking methodology and its underlying assumptions. Worse still, essential details
of many of the algorithms currently used by the BCS are not even
openly declared (the only one
completely declared in the public domain is the system by Colley
\mcite{ColleyWeb}, though some others are at least partially explained). 
Of those that are at least partially public,
some include seemingly arbitrary
parameters whose effects are difficult to interpret, and others are
tweaked periodically to obtain the purportedly most reasonable
ranking based on previous results.

In this article, we demonstrate that a simply-explained algorithm
constructed by crudely mimicking the behavior of voters can provide
reasonable rankings. We define a collection of voting automatons (random walkers) each of whom declares its preference for a single team. Each automaton repeatedly selects a game at random from its preferred team's schedule and decides whether to change its preference to the opponent as biased by the game outcome, preferring but not absolutely certain to go with
the winner, repeating this process indefinitely. In the simplest implementation of this process, the
probability $p$ of choosing the winner is kept constant across voters
and games played, with $p > 1/2$ because on average the winner should
be recognized as the better team and $p < 1$ to allow a given voter to argue that
the losing team is still the better team (moreover, the $p = 1$
limit can be mathematically more complicated in certain scenarios, as
discussed in section \ref{sec:asymptotics}).
The voting automatons are nothing more than
independent, biased random walkers on a graph connecting the teams (vertices) by their head-to-head games (edges).
These ``voters'' thereby obey idealized behavioral rules dictated by one of the most natural arguments relating the relative ranking of two teams: ``my team beat your team.'' Indeed, the statistics of such biased random walkers can be presented as nothing more than the logical extension of this argument repeated \emph{ad infinitum}.

This algorithm is easy to explain in terms of the
``microscopic'' behavior of individual walkers who randomly change
their opinion about which team is best (biased by the outcomes of individual games).  Of course,
this behavior is grossly simplistic compared with real-world 
poll voters. In fact, under the
specified range of $p$, a single walker will never reach a definitive
conclusion about which team is the best; rather, it will forever
change its allegiance from one team to another, ultimately traversing
the entire graph. However, the ``macroscopic'' total
number of votes cast for each team by an aggregate of random-walking
voters quickly reaches a statistically-steady ranking.

The advantage of the algorithm discussed here is that it can be easily
understood in terms of single-voter behavior. Additionally, it has a
single explicit, precisely-defined parameter with a meaningful
interpretation at the single-voter level.  We do not claim that this
ranking is superior to other algorithms, nor do we review the vast
number of ranking systems available, as numerous reviews are already available
(see, for example, \mcite{Keener93}, \mcite{Stefani97}, 
\mcite{ConnorGrant00}, \mcite{Martinich02} for reviews of
different ranking methodologies and the list of algorithms and
``Bibliography on College
Football Ranking Systems'' maintained by \mcite{dwilson}).  We do
not even claim that this ranking algorithm is wholly novel; indeed,
the resulting linear algebra problem is in the class of ``direct
methods'' discussed by Keener \mcite{Keener93} and has many similarities to
the linear algebra problem solved by Colley \mcite{ColleyWeb}.  Rather,
we propose this random-walker ranking on the strength of its simple
interpretation: our intent is to show that this simply-defined
ranking yields reasonable results.

The remainder of this article is organized as follows: In section
\ref{sec:ranking}, we give the mathematical definition of the ranking
algorithm and examine its statistical properties.  
In section \ref{sec:asymptotics}, we
investigate the algorithm's asymptotic behavior for extreme values of
the probability $p$. In section \ref{sec:roundrobin}, we present results for special cases involving round-robin tournaments.  We then
examine recent historical outcomes from real NCAA Division I-A seasons
in section \ref{sec:recent}. In section \ref{sec:network}, we discuss
properties of the graphs defined by the games played in a given year
according to real NCAA Division I-A schedules, paying special
attention to hierarchical structure and the interplay between that
structure and random-walker rankings. We conclude in section \ref{sec:summary} by discussing some possible generalizations of this ranking algorithm.

\section{RANKING WITH RANDOM WALKERS.}   
\label{sec:ranking}

For each team $i$ we denote the number of games it played by $n_i$, its
wins by $w_i$, and its losses by $l_i$. A tie, which was possible prior to the
current NCAA overtime format, is treated as half a win and half a
loss, so $n_i=w_i+l_i$ always holds.  The number of random walkers
declaring their preference for team $i$ is denoted $v_i$, and the
condition that the total number of voters remain constant is given by $\sum_i
v_i = Q$.

If team $i$ beats team $j$, then the average rate at which a walker
voting for team $j$ changes its allegiance to team $i$ is proportional
to $p$, and the rate at which a walker already voting for team $i$
switches to team $j$ is proportional to $1-p$.  For simplicity, we
ignore the dates of games, selecting each of the $n_i$ games played by
team $i$ with equal probability.  To avoid rewarding teams for playing
more games, the rate at which a given voter considers a given game is
taken to be independent of $n_i$.  Under this selection of rates, we
find it more natural to express the independent random-walker dynamics
using ordinary differential equations (ODEs) instead of Markov chains, though our entire
discussion can certainly be recast in terms of Markov chains, in which case 
the approach has some similarities with the PageRank citation ranking
\mcite{BrinPage99} underlying the Google search engine (see 
\mcite{LangvilleMeyer05}). 
Formulating the present problem in terms of ODEs has the added advantage of avoiding complications due to cycles of different lengths (such cycles could alternatively be removed through
explicit time-averaging).

The expected rate of change of the number of votes cast for each team
in this random walk is quantified by a homogeneous system
of linear differential equations,
\begin{equation}
  \mathbf{{v}}' = \mathbf{D}\mathbf{{v}}\,,\label{eqn:D}
\end{equation}
where $\mathbf{{v}}$ is the $T$-element vector of
the number ${v}_i$ of votes cast for each of the 
$T$ teams, and the elements of the square matrix $\mathbf{D}$ are
\begin{align}
        D_{ii} &= -pl_{i}-(1-p)w_{i}\,,
\notag \\
        D_{ij} &= \frac{1}{2}N_{ij} + \frac{(2p-1)}{2}A_{ij} 
        \quad (i\neq j)\,, \label{eqn:Ddef}
\end{align}
in which $N_{ij}$ is the number of games played between
teams $i$ and $j$ and $A_{ij}$ is the number of times team $i$ beats
team $j$ minus the number of times $i$ loses to $j$. That is, 
if $N_{ij}$ belongs to $\{0,1\}$, then
\begin{equation}
A_{ij} = \left\{\begin{array}{cc}
+1 & \mbox{if team } i \mbox{ beat team } j\,,\\
-1 & \mbox{if team } i \mbox{ beat team } j\,,\\
0  & \mbox{if team } i \mbox{ beat team } j\,.\end{array}
\right.
\end{equation}
When two teams play each other multiple times,
we obtain $A_{ij}$ as summed over those games.  Because $1/2<p<1$, the off-diagonal 
elements $D_{ij}$ are nonnegative and vanish if and only if $N_{ij} = 0$.

Ideally, one would consider the giant connected component in the
college football schedule network, including 
Division I-A as a subgraph, but it was easier to analyze the data (obtained from
\mcite{Howell} and \mcite{Wolfe}) when the information was selectively restricted to the graph of
Division I-A teams. Because many Division I-A
teams play some non-I-A opponents, we 
represent all of these connected non-I-A teams collectively as a single node. 
These teams usually do not fare well against Division I-A
competition, so this new ``team'' 
achieves a low ranking and does not
significantly affect the random walker populations, except to penalize
the Division I-A teams they defeat and to maintain the constraint that
the total number of votes $Q$ remain constant (i.e., voters do not
leave the graph).

\bigskip\noindent
{\bf Equilibrium.}
The matrix $\mathbf{D}$ encompasses all the connections and win-loss 
outcomes between teams. The steady-state equilibrium of (\ref{eqn:D},\ref{eqn:Ddef}) satisfies
\begin{equation}
        \mathbf{D}\mathbf{\bar{v}} = \mathbf{0}
  \label{eqn:Dsteady}  
\end{equation}
and gives the expected populations
$\mathbf{\bar{v}}$ of the random walkers voting for each team.  
This information can then be used directly to rank the 
teams. Despite the simplistic behavior of an individual
random walker, the behavior of an aggregate of voters (or
equivalently, because of the independence assumption, the long-time
average of a single voter) appears to yield reasonably robust orderings
of the top teams.  Unsurprisingly, the 
number of votes cast for a given team varies substantially for different 
values of $p$; nevertheless, the relative ranking of the top few teams can
remain similar across a wide range of $p$, as we discuss for recent
historical examples in section \ref{sec:recent}.

The equilibrium point $\mathbf{\bar{v}}$ lies in the null-space of
$\mathbf{D}$; that is, it is the eigenvector associated with a zero
eigenvalue.  We stress that this equilibrium does not require a no-net-flow
detailed balance along each edge; rather, we only require zero
net flow out of each node. For instance, a schedule
of three teams ($i$, $j$, and $k$) in which each plays only two games such that $i$ beats
$j$, $j$ beats $k$, and $k$ beats $j$ leads to a cyclic flow of votes
around the triangle with a statistical
equilibrium that arises when each team receives an equal number of votes.

An important property of these random walkers is that the matrix
$\mathbf{D}$ yields a single attracting equilibrium
$\mathbf{\bar{v}}$ for a given $p$ in $[1/2,1)$, provided the underlying
graph representing games played between teams consists of a single connected
component. This can be proved in three steps:
\begin{enumerate}
\item{The column sums of $\mathbf{D}$ vanish because the sum of the
    populations remains constant, $0=Q'=\sum_i v_i'=\sum_i\sum_j
    D_{ij}v_j$ (i.e., the dynamics of the ${v}_j$ are confined to a
    hyperplane of codimension $1$).}
\item{The off-diagonal elements of $\mathbf{D}$ are nonnegative and---once 
    the graph consists of a single connected component---all
    off-diagonal elements of $D^d$ are positive, where $d$ is the diameter
    of the graph, so that vertices with ${v}_j=0$ have growing 
    populations. In other words, all average flows enter the 
    hyperquadrant in which ${v}_j > 0$ for all $j$.}
\item{Finally, because (\ref{eqn:D}) is linear, the only possibility
consistent with the foregoing observations is that there is a single
attracting sink $\bar{v}$ in the hyperplane.}
\end{enumerate}
This argument breaks down if either the graph is not connected or
$p=1$, because subgraphs with zero walker population can
remain so. Alternatively, one can recast 
(\ref{eqn:D})--(\ref{eqn:Dsteady}) as an eigenvalue problem and apply the
Perron-Frobenius theorem, as described by \mcite{Keener93}.  However, given
the random-walker interpretation built into the rate matrices
we define, our arguments already ensure that the expected populations
achieve a unique attracting state, provided the
graph consists of a single connected component and $p<1$. In the absence of these conditions, the Perron-Frobenius theorem cannot be applied 
because the resulting matrices are no longer irreducible.

\bigskip\noindent
{\bf Statistics.}
Any initial voter distribution eventually randomizes completely, so
the steady-state distribution of the number of votes $\bar{v}_i$ cast for
the $i$th team is binomial (for $Q$ trials) with probability
$\bar{v}_i/Q$, mean $\bar{v}_i$, and variance
$\bar{v}_i(1-\bar{v}_i/Q)$.  The joint probability density
function of two vertices is not perfectly independent, as the sum over
all vertices equals the number of random walkers $Q$.  However, it is
still obtained from a binomial distribution of $Q$ random trials across the
vertices.  (The two of interest have probabilities $r_i=\bar{v}_i/Q$ and 
$r_j=\bar{v}_j/Q$.) We can exploit this fact to measure the
confidence in the relative ranking of two teams in terms of the
minimum number of voters $Q_{min}$ required to ensure that the
expected difference between the number of votes cast for each team is
larger than the standard deviation of that difference:
\begin{equation}
  \label{eqn:Qmin}
  Q_{min} = \frac{r_i+r_j-(r_i-r_j)^2}{(r_i-r_j)^2} 
  = \frac{r_i+r_j}{(r_i-r_j)^2} - 1\,.
\end{equation}
Because the statistical properties of the random walkers
follow directly from the linear algebra problem
(\ref{eqn:D})--(\ref{eqn:Dsteady}), there is no need to
simulate independent random walkers to obtain rankings.
This simplicity disappears if interactions between random walkers are
included, as considered briefly in section \ref{sec:summary}.

\section{ASYMPTOTICS AT LARGE AND SMALL $p$.}
\label{sec:asymptotics}

For a given probability $p$ the
expected populations depend in a complex manner on the details of game
schedules and outcomes. In an attempt to clarify the effects
of selecting a given $p$, it is instructive to 
investigate analytically 
the limiting behaviors near $p = 1/2$ and $p = 1$. We
demonstrate, in particular, that the main contributions near $p = 1/2$
include a measurement of schedule strength, whereas behavior near $p =
1$ is dominated by undefeated teams and by subgraphs of teams that go
undefeated against teams outside the subgraph.

\bigskip\noindent
{\bf On any given Saturday.}
Consider $p=1/2+\varepsilon$, where $\varepsilon\ll 1$.  The rate matrix 
(\ref{eqn:Ddef}) becomes
\[
D_{ij} = \frac{1}{2}\Delta_{ij} + \varepsilon\widetilde{D}_{ij}\,,
\]
where $\mathbf{\Delta}$ is the graph Laplacian: 
$\Delta_{ij}=1$ ($i\neq j$) if nodes $i$ and $j$ are connected (the
two teams played each other) and $\Delta_{ii}=-n_i$.
By (\ref{eqn:Ddef}), ${\mathbf{\widetilde{D}}}$ has the same elements as
$\mathbf{A}$ off the diagonal and the values $w_i-l_i$ on the
diagonal. A power series of the equilibrium probabilities,
$\bar{v}_j=\bar{v}_j^{(0)}+ \varepsilon\bar{v}_j^{(1)} +
\varepsilon^2\bar{v}_j^{(2)} + \cdots$, then satisfies
\begin{equation}
  \label{eq:orders}
  0 = \frac{1}{2}\mathbf{\Delta\bar{v}}^{(0)} + 
  \sum_{k=1}^\infty \varepsilon^k\left[
    \frac{1}{2}\mathbf{\Delta\bar{v}}^{(k)} +
    \mathbf{\widetilde{D}\bar{v}}^{(k-1)}\right]\,,
\end{equation}
subject to the normalization condition $\sum_j
\bar{v}^{(k)}_j=Q\delta_{k0}$ for $Q$ voters ($\delta_{k0}=1$ when $k=0$, otherwise $\delta_{k0}=0$).

The $O(\varepsilon^0)$ contribution requires $\bar{v}_j^{(0)}=Q/T$ for
each $j$, distributing the $Q$ votes equally across the $T$ nodes of
the graph.  With $\widetilde{D}_{ij}\bar{v}_j^{(0)} = 2(w_i-l_i)Q/T$, the
$O(\varepsilon^1)$ condition becomes
\begin{equation}
  \label{eq:order1}
  \sum_j\Delta_{ij}\bar{v}_j^{(1)} = -\frac{4Q}{T}(w_i-l_i)\,.    
\end{equation}
That is, the first correction $\bar{v}_j^{(1)}$ at $p$ near $1/2$
is a potential that satisfies a discrete Poisson equation (subject to
the constraint $\sum_j\bar{v}_j^{(1)}=0$) with charges proportional to
the win-loss record of each team.  It thus incorporates the record
of the $j$th team and is also heavily influenced by the records of the
nearest neighbors and other close teams in the graph. In other words, the first
correction is strongly influenced by a ``strength of schedule'' notion. 
Furthermore, it is only with the second-order 
term $\bar{v}_j^{(2)}$ that information pertaining to specific games won or lost by a given 
team begins to be incorporated, as $\bar{v}_j^{(1)}$ includes 
only net records in (\ref{eq:order1}).

\bigskip\noindent
{\bf Winner takes all.}
The asymptotic behavior for $p=1-\varepsilon$ ($\varepsilon\ll 1$) is more complicated because the
limiting state depends on the number of undefeated teams and
other schedule details. The single-equilibrium argument of section
\ref{sec:ranking} breaks down at $p=1$ because off-diagonal
elements of $D^d$ are not necessarily positive. There can then be
multiple equilibrium states, but only one of these states
is achieved in the limit as $p\to 1$.

The simplest situation to consider asymptotically occurs when a single
undefeated team garners all random walker votes for $p=1-\varepsilon$
as $\varepsilon\to 0$. This condition requires both that there be only
a single undefeated and untied team and that there be no subgraphs
of other teams in the network that collectively win all of their games
against teams outside that subgraph. 
The transition rates are then written $D_{ij} =
D_{ij}^{(0)}-\varepsilon\widetilde{D}_{ij}$, and we again expand
\[
\bar{v}_j=\bar{v}_j^{(0)}+ \varepsilon\bar{v}_j^{(1)} +
\varepsilon^2\bar{v}_j^{(2)} + \cdots.
\]
Here, $\mathbf{D}^{(0)}$
agrees with $(\mathbf{N}+\mathbf{A})/2$ off the diagonal but has the negation
of the number of losses by the corresponding teams on the diagonal
(recall that a tie counts as half a win and half a loss), and
$\mathbf{\widetilde{D}}$ remains as defined earlier in the perturbation around $p=1/2$.
The limiting state $\bar{v}_j^{(0)}=Q\delta_{ju}$ casts all $Q$ votes for the
single undefeated team (``$u$''). Calculation of the first-order correction 
then requires solution of
$\mathbf{D}^{(0)}\mathbf{\bar{v}}^{(1)} = \mathbf{b}$, where
$b_j=Q\widetilde{D}_{ju}$, which offers neither simplification nor intuition
beyond the original rate equations
(\ref{eqn:D})--(\ref{eqn:Dsteady}).

\section{ROUND-ROBIN EXAMPLES.}   
\label{sec:roundrobin}

The asymptotic analyses of the previous section were necessarily
limited by the generality of possible schedule topologies.  As a
means of developing further intuition, we consider the special
cases of round-robin tournaments in which every team plays every other
team exactly once.

\bigskip\noindent
{\bf On any given Saturday (revisited).}
Returning to the case $p=1/2 + \varepsilon$, consider a round-robin tournament of $T$
teams, each of which plays exactly $T-1$ games, one against each of the available
opponents.  Then the graph Laplacian is the matrix with $1$s off the
diagonal and $-(T-1)$ for each diagonal entry. Subject to the
constraint $\sum_j\bar{v}_j^{(k)}=0$ ($k>0$), we immediately
obtain $\sum_j\Delta_{ij}\bar{v}_j^{(k)} = -T\bar{v}_i^{(k)}$, so
(\ref{eq:orders}) yields
\[
\bar{v}_i^{(k+1)} = \frac{2}{T}\widetilde{D}\bar{v}_i^{(k)}\,.
\]
The zero-sum constraint is maintained because the column sums of
$\widetilde{D}$ are identically zero by definition. This simple iterative
relation for $\bar{v}_i^{(k+1)}$ can then be summed when 
$\varepsilon<1/2$
(i.e., $p<1$) to arrive at
\[
\bar{v} = \left[ I - \frac{2\epsilon}{T}\widetilde{D}\right]^{-1} \bar{v}^{(0)}\,.
\]

We make special note of the first-order correction. In
section \ref{sec:asymptotics}, we saw that strength of schedule and
the direct win-loss record both appear at $O(\varepsilon)$. Hence, one
expects the win-loss record of an individual team to
determine its $O(\varepsilon)$ contribution fully in the round-robin case,
where everyone playing everyone else removes schedule inequities.
Indeed, equation (\ref{eq:order1}) gives
\begin{equation}
  \label{eq:order1rr}
  \bar{v}_i^{(1)} = \frac{4Q}{T^2}(w_i-l_i)
\end{equation}
when applied to round-robin tournaments.  Stated differently, the asymptotic rankings near $p=1/2$ in a round-robin tournament are linear, and the slopes
are set by the win-loss records.

\bigskip\noindent
{\bf Perfectly ordered teams.}
Because the asymptotic expansion for $p=1-\epsilon$ is complicated by the
difficult determination of the base state ($\epsilon=0$), we consider
this expansion only for the round-robin case of $T$ ``perfectly
ordered'' teams, which we define as the special case where team $i$
beats team $j$ whenever $i<j$. In this situation, the $\mathbf{D}^{(0)}$ matrix
of section \ref{sec:asymptotics} is upper triangular, with $1$s above the
diagonal and $D_{ii}=-l_i=-(i-1)$. Because there are no cycles here, it is clear
that the single undefeated team garners all votes in
the limit as $\varepsilon\to 0$, so $\bar{v}_i^{(0)} = Q\delta_{i1}$.  
We can
then show that
\[
\bar{v}_i^{(1)} = \left\{
    \begin{array}{cc}
      -(T-1) & \mbox{if } i=1\,, \\
      T/[l_i(l_i+1)] & \mbox{if } i>1\,.
    \end{array}\right.
\]

\bigskip\noindent
{\bf Mixed ordering.}
Finally, it is worth asking whether round-robin tournaments can
give any indication about useful values of the bias parameter
$p$.  Specifically, it is reasonable to expect that teams in a
round-robin schedule should be ranked in a manner consistent
with their win-loss records, as their schedules contain no inequities.
A 6-5 team in a 12-team round robin should presumably be ranked higher
than a 5-6 team, even if only marginally so. As we already saw from
(\ref{eq:order1rr}), the $O(\varepsilon)$ term in an
expansion with $p=1/2+\varepsilon$ for a round-robin competition  does indeed agree
with the rank ordering given by win-loss records.

Do the terms that are higher order in $\varepsilon$ continue to respect
the win-loss ordering in a round-robin setting?  Equivalently, what happens for values of $p$ further
from $1/2$? One might reasonably ask whether this win-loss ordering is
generically preserved for all $p$ in the range $1/2<p<1$ (it is not). It is then
natural to ask whether the win-loss ordering is always preserved up to
``crossing'' values bounded away from $1/2$. That is, for a given
round-robin outcome, we define $p_{c}$ to be the minimum 
$p$ ($\geq 1/2$) such
that the random-walker ordering crosses from consistent to inconsistent with win-loss records.
We then ask whether $p_{c}$ is bounded from below.

However, it seems clear from examples of specific outcomes that there
is no lower bound for $p_{c}$ except for $p=1/2$ (where, by definition,
all teams are ranked equally).  
We obtain $p_{c}$ arbitrarily close to $1/2$ through a simple
modification to the perfectly-ordered tournament.
Starting from a win-loss matrix of $T$
teams where team $i$ beats $j$ if $i<j$, we modify only the games
played by the team with the best losing record. To be precise, for even $T$
we change only games played by team $k=(T/2)+1$, with $(T/2)-1$ wins
and $T/2$ losses. We now swap the win-loss outcome of every game
played by team $k$ except for the game between $k$ and $k-1$, which we
maintain as a win for team $k-1$. These outcome switches modify the win-loss record of every team except for teams $k$ and
$k-1$; in particular, the latter still has a better record than the
former. Nevertheless, we observe numerically 
that $(p_{c}-1/2) \sim T^{-1}$ for these
explicitly-constructed mixed-ordered round-robin events,
with team $k$ ranked above team $k-1$ above $p_{c}$.

The realization that there is no lower bound for $p_{c}$ away from
$1/2$ indicates an obvious
limitation to this simply-constructed ranking system.
Specifically, because the random walkers inherently represent
first-place votes, rank-ordered crossings involving teams ranked
\#$T/2$ and \#$(T/2)+1$ are not very surprising.  The number of votes
cast for each team in the statistical equilibrium is of course
important, in that the few votes cast for the lowest-ranked teams by
design impact the number of votes cast for each team they play, ensuring
that strength of schedule is inherently incorporated.  However, the emphasis
on highly-ranked teams means that a team can improve its ranking by
beating a highly-ranked team more than it might be penalized for losing
to a lower-ranked team. These round-robin examples then clearly call into question the accuracy of the
precise rankings of those middle teams and thus lead to questions
about the scheme as a whole.  A system based only on
first-place votes should presumably be strongest in its ranking of the
top teams, though we have no mathematical proof
that these walkers do so in any ``optimal'' sense. This motivates us to
consider how the random-walker rankings fare in comparison with the historical
record.

\section{RECENT HISTORY.}    
\label{sec:recent}

The utility of a ranking algorithm lies in its performance in the face
of real data.  Accordingly, we investigated college football rankings
for each season from 1970 to 2005, restricting our discussion
here to 2001 and 2002, as these seasons represent recent extreme
situations in attempting to pick the top two teams (see, for example,
\mcite{BCS}). The 2003, 2004, and 2005 seasons are discussed on our 
web page \mcite{ourweb}, where rankings for future seasons will also 
be maintained.

\begin{figure}
  {\psfragscanon\footnotesize
    \psfrag{p}{$p$}
    \psfrag{2001 BCS}{\hspace*{0.05in}2001 pre-bowl}
    \psfrag{2002 BCS}{\hspace*{0.05in}2002 pre-bowl}
    \psfrag{Rank}{Rank}
    \centerline{ (a)\hspace{-0.15cm}
      \includegraphics[width=0.45\textwidth]{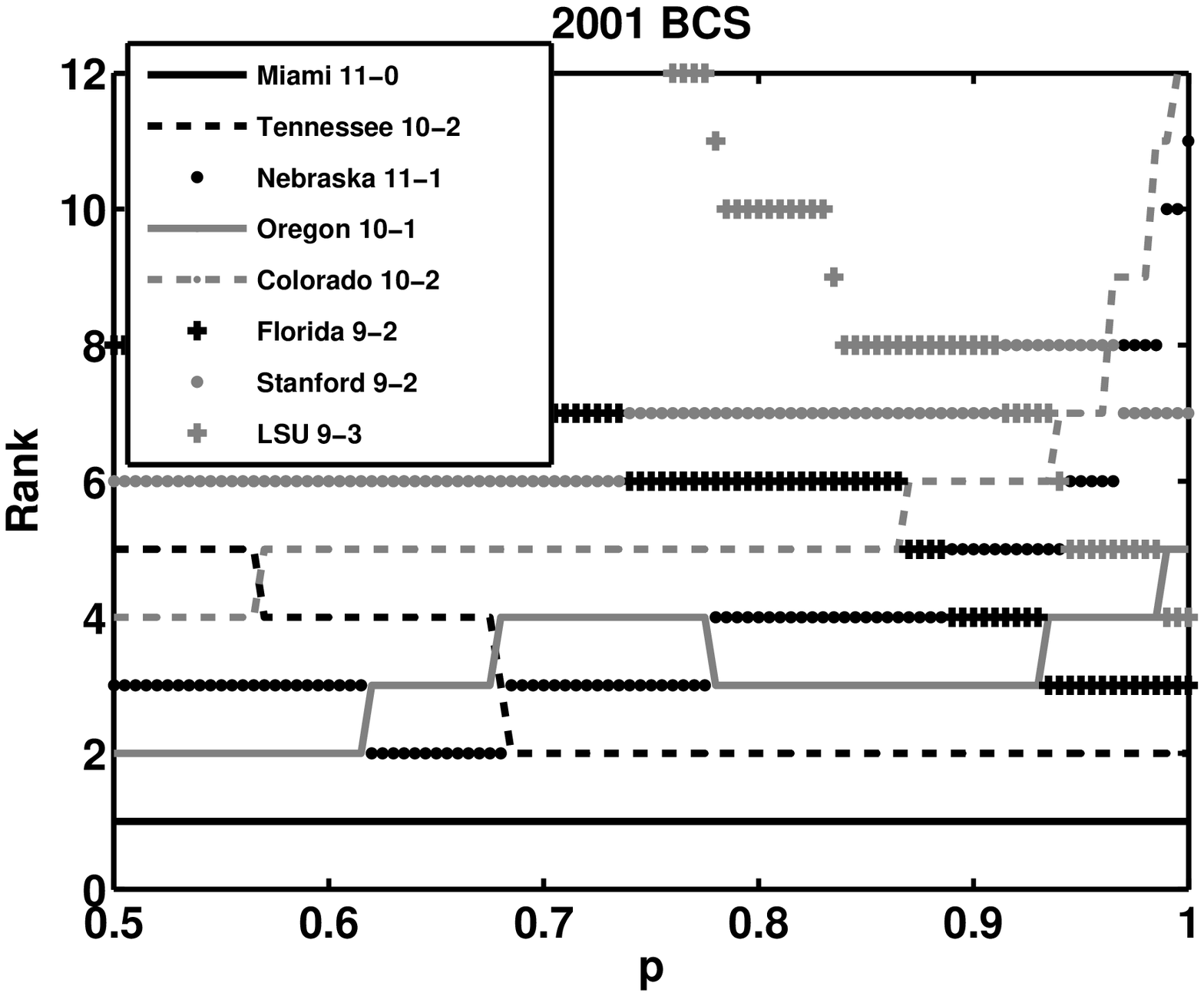}
      \hspace{0.2cm} (b)\hspace{-0.15cm}
      \includegraphics[width=0.46\textwidth]{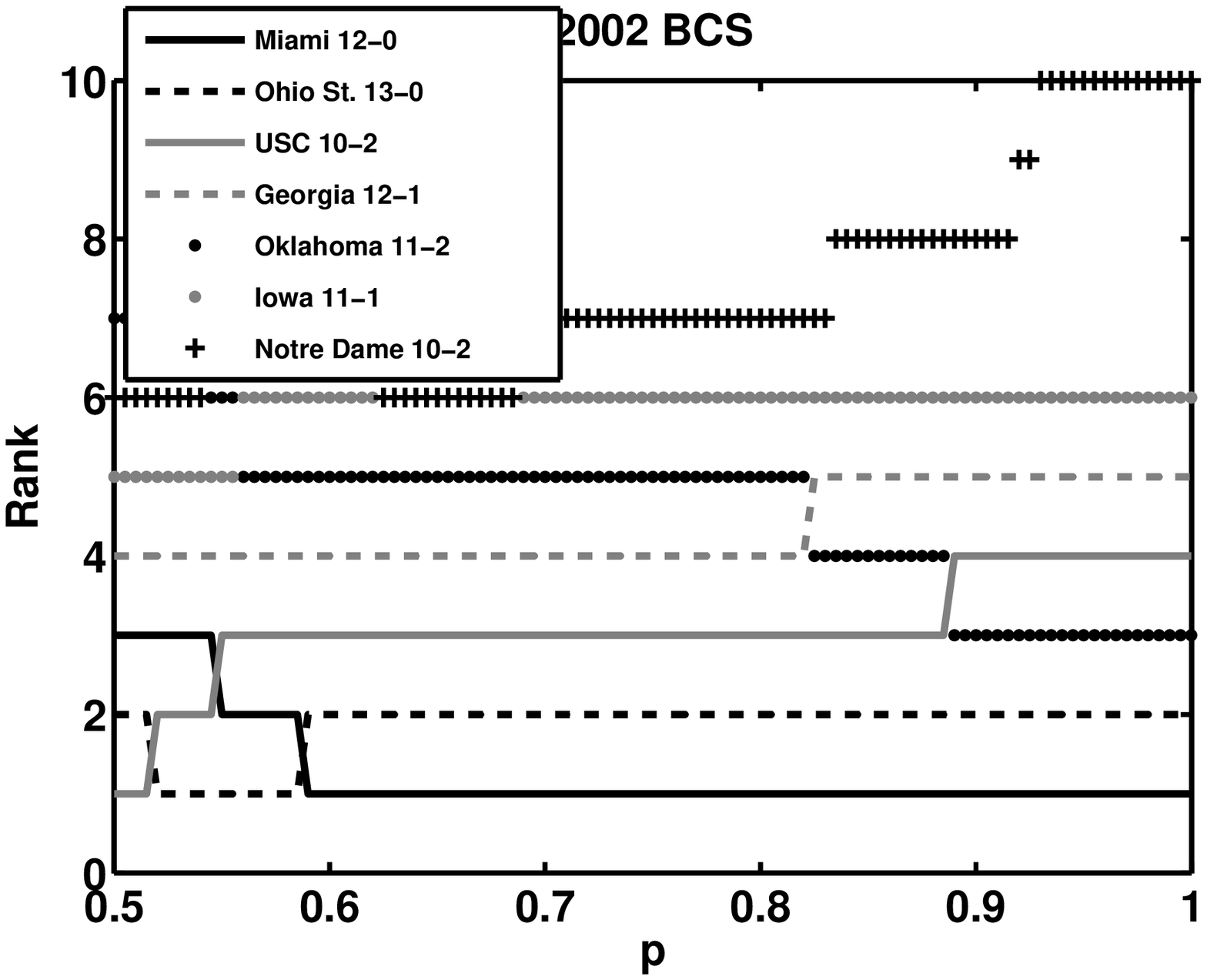}
    }
  }
  \caption{Pre-bowl rank orderings of teams by the expected
    populations of random walkers at different values of
    the probability $p$ of voting for the winner of a given game
    for (a) 2001 and (b) 2002.}
  \label{fig:populations}
\end{figure}

The only certainty in the 2001 pre-bowl rankings
was that Miami belonged in the National Championship Game, because it
was the only undefeated team in Division I-A. Indeed, both polls and
all eight algorithms used by the BCS that year picked Miami \#1 going
into the bowl games. That season's controversy concerned Nebraska's
selection as the \#2 BCS team, narrowly surpassing BCS \#3 Colorado
despite Colorado's late-season rout of Nebraska.  The fact that BCS
\#4 Oregon had been ranked \#2 in both polls was also mentioned on
occasion.  After the bowl games, in which Miami defeated Nebraska and
Oregon defeated Colorado, it was Oregon's absence from the
championship game that became the centerpiece of national controversy.
The random walkers select Oregon \#2 for $p>.5$ up to $p \approx
.62$, above which Nebraska takes \#2 in a narrow range up to $p\approx
.68$. Above that value, the random walkers defy conventional wisdom by
selecting BCS \#6 Tennessee as \#2.  In fact, they choose Tennessee as \#2
over the widest range of probabilities (see Figure
\ref{fig:populations}a).  This ranking is explained in part by
the fact that the simplest random-walking voter algorithm presented
here does not distinguish games based on the date played. Tennessee
had been the most likely team to be picked to face Miami until they
lost the Southeastern Conference (SEC) championship game on the final day before the BCS
selections.  Even with the loss (because the date of that loss does not affect things here), the random walkers congregate near Tennessee,
in part because the SEC as a whole is highly ranked that season.  Indeed, in
the limit as $p\to 1$, Florida is ranked \#3 and LSU is ranked \#4, with Oregon falling to \#5 and both
Nebraska and Colorado falling completely out of the top ten. This SEC near-dominance is generated by a number of nearly-undefeated subgraphs: Tennessee (10-2) lost to Georgia and lost to LSU the second time they played (in the SEC championship), Florida (9-2) lost to Auburn and Tennessee, and LSU (9-3) lost to Mississippi, Florida, and to Tennessee in their first meeting. Not counting games against each other, this trio of closely-linked teams had a collective 24-3 record prior to the bowl games, with all three losses against SEC conference opponents, two of whom were played multiple times by this group (Florida beat Georgia and LSU beat Auburn).  Those three losses came against teams who also had all but one of their losses in the SEC (Auburn lost to Syracuse).  Florida falls to \#8 at smaller values of $p$ (closer to $1/2$), and LSU is not even in the top ten for $p$ below $\approx .77$.  Nevertheless, they continue to help Tennessee's ranking.

In contrast, the BCS system worked virtually without controversy in
selecting teams for the National Championship Game at the end of the
2002 season, as there were precisely two undefeated
teams, both from major conferences.  
Both polls picked Miami and Ohio State as the top two
teams, in agreement with six of the seven ranking algorithms used that
year.  The only nonconformist was the \emph{New York Times} ranking system,
which picked Miami and USC as the top two teams---the latter
presumably in part due to its difficult schedule. For $p \approx 1/2$,
the random walkers also rank USC in the top two based on the strength
of its schedule (see Figure
\ref{fig:populations}b), but they agree across
most values of $p$ that the top two teams are Miami and Ohio State.

\begin{figure}
  {\psfragscanon\footnotesize
    \psfrag{p}{$p$}
    \psfrag{2001 BCS}{\hspace*{-0.15in}2001 pre-bowl}
    \psfrag{2002 BCS}{\hspace*{0.55in}2002 pre-bowl}
    \psfrag{1/Q}[b][b]{$1/Q_\mathrm{min}$}
    \centerline{(a)
      \includegraphics[width=0.45\textwidth]{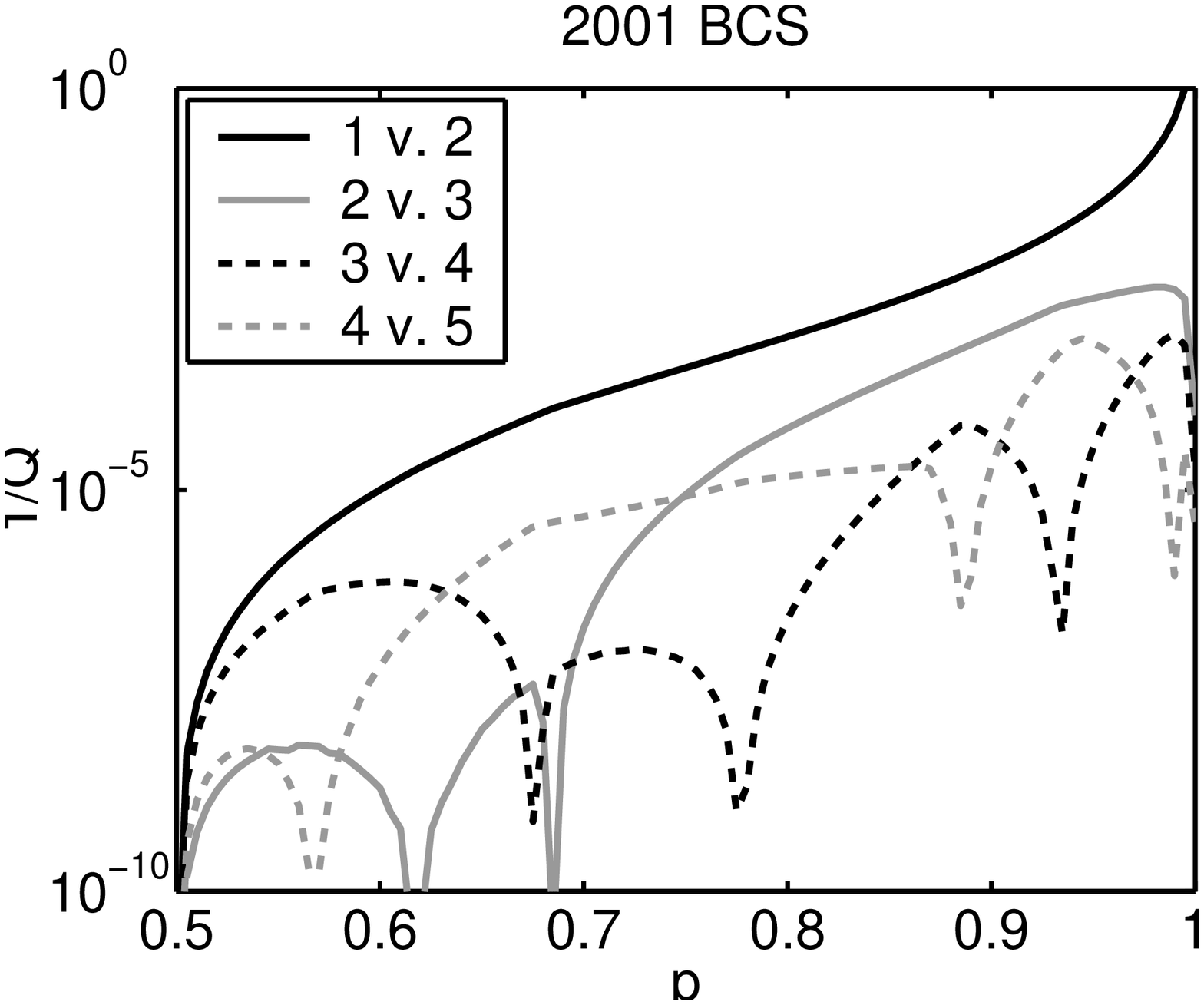}
      \hspace{0.1 cm} (b)
      \includegraphics[width=0.45\textwidth]{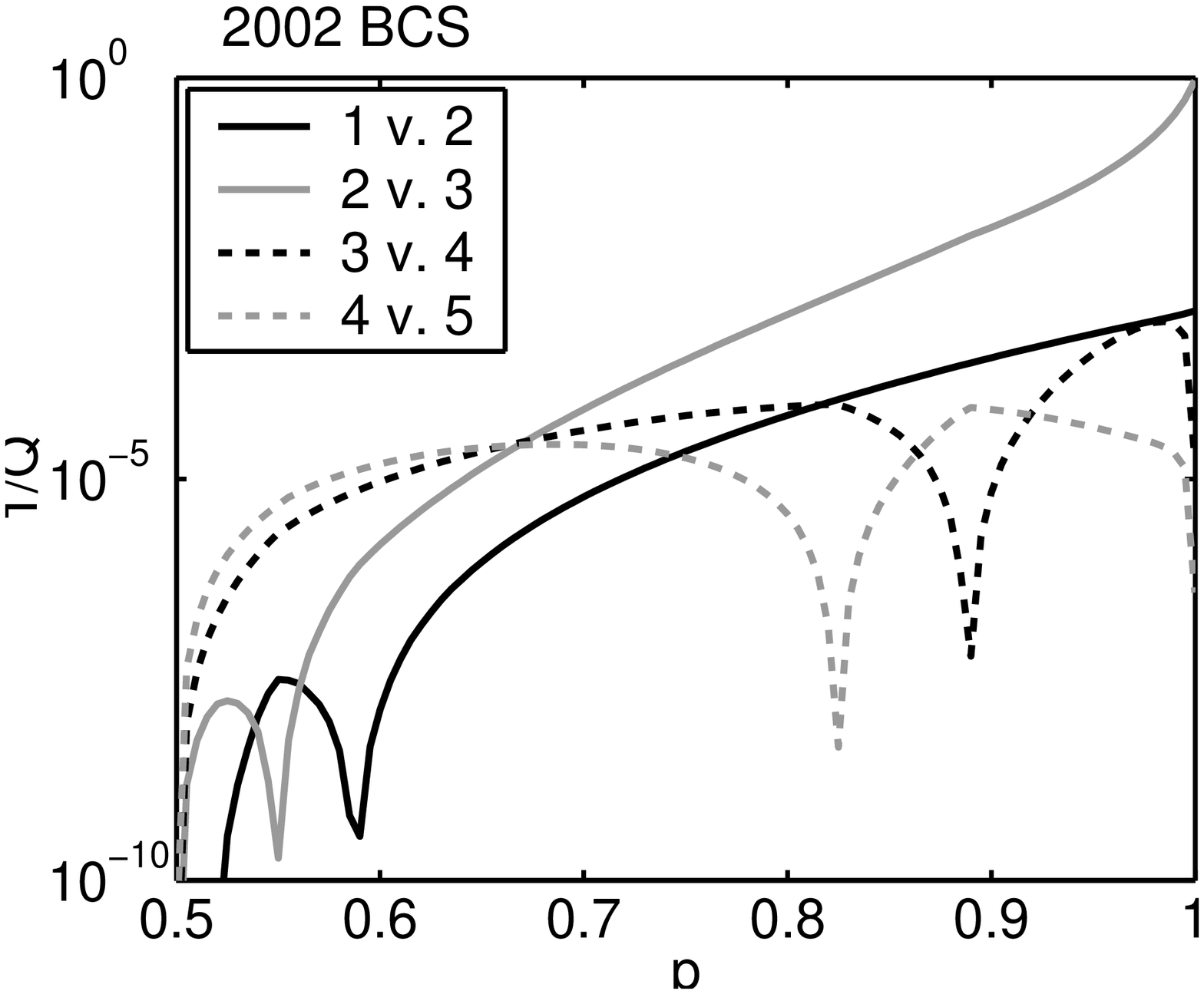}
    }
  }
  \caption{Degree of confidence in the ordering of pairs of teams in the
    (a) 2001 and (b) 2002 pre-bowl rankings, as quantified
    by $1/Q_{min}$, where $Q_{min}$ denotes the minimum number of independent
    random walkers necessary to ensure that the standard deviation of
    the difference between the expected populations of two teams is
    smaller than the expected difference (see equation (\ref{eqn:Qmin})).}
  \label{fig:fluctuations}
\end{figure}

One can explore the differences between the 2001 and 2002 pre-bowl results further by exploiting the statistical properties of expected
vote totals to express a measure of confidence in the resulting
rankings. The $Q_{min}$ required to distinguish successfully between
the top rank-ordered pairs of teams, defined by (\ref{eqn:Qmin}) in section
\ref{sec:ranking}, is plotted for the 2001 and 2002 pre-bowl rankings
in Figure \ref{fig:fluctuations}. In particular, we note that the
relative numbers of voters required to distinguish \#1
from \#2 in 2001 and \#2 from \#3 in 2002 are significantly smaller
(indicating a higher degree of confidence in the chosen ordering) than, in particular, the
distinctions between \#2 and \#3 in 2001 or \#1 and \#2 in 2002.
One should, of course, be careful in applying this measure of confidence too broadly, because the percentage of total votes that remain available to distinguish \#4 from \#5 will in general be smaller than those available to distinguish \#1 from \#2. Accordingly, direct comparisons of the same rank distinction across different years are more reasonable.  For example, observe the typically lower degree of confidence in distinguishing \#3 from \#4 and \#4 from \#5 in 2001 (when Nebraska, Colorado, Oregon, and Tennessee were all trying to lay claim to the \#2 spot) compared with the situation in 2002.

\begin{figure}[ht]
  {\psfragscanon\footnotesize
    \psfrag{X pop}[t][t]{Population of Team X}
    \psfrag{Y pop}[t][t]{Population of Team Y}
    \centerline{
      \includegraphics[width=0.8\textwidth]{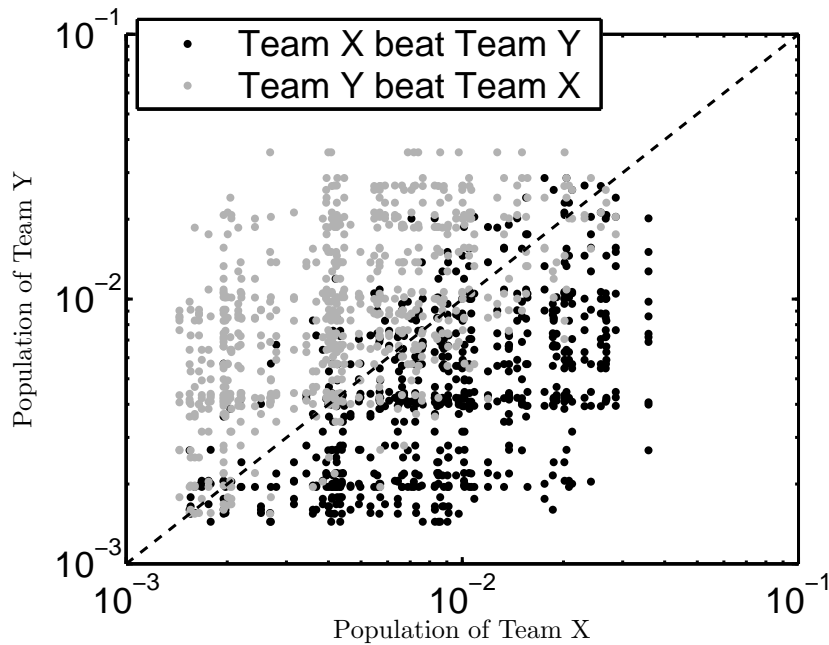}
    }
  }
  \caption{Wins and losses plotted according to the 2001 pre-bowl 
    rankings of each team at p=0.8, given by the expected fraction of
    random walkers populating (``voting for") each team.}
  \label{fig:violations}
\end{figure}

After seeing these results, we should discuss at least briefly which values of $p$ might yield good rankings. Perhaps
neither the dominance of strength of schedule near $p=1/2$ (as illustrated, for example, by USC in 2001) nor the emphasis on undefeated seasons near
$p=1$ (e.g., BYU in 1984) are appropriate. Rather, we might
naively argue that a balanced value of $p$ is preferred.

Alternatively, we could consider various objectives for 
optimizing $p$. For instance, Figure \ref{fig:violations}
shows the organization of the win-loss outcome of every game in the
2001 pre-bowl schedule according to the random walker populations determined
with $p=0.8$. The votes are not (of course) strictly ordered according to the winner
and loser of each individual game, but the winners have
more votes on average. Instances where a higher-ranked team has lost to a lower-ranked team are frequently referred to as ``ranking violations'' and can be used
to construct one measure of the quality of different ranking systems (see the comparisons page maintained by Massey \mcite{MasseyComparisons}).  Additionally, the goal of minimizing the number of such violations and determining rank orderings from the multitude of such nonunique minimizations can itself be used to define ranking algorithms \mcite{Coleman05}, \mcite{Park05}.  An obvious optimization procedure here is to
select $p$ to minimize the number of ranking violations. While we observe that this optimal $p$ varies from year to year, it frequently takes a value near the center of the allowed interval.  Other measures of the quality of the ranking may be naturally inspired by Figure \ref{fig:violations}; for example, we might minimize the violations in another norm, such as one giving their distance from nonviolation.
Another approach based on logistic regression of a generalized version of the present ranking system has been implemented by Kvam and Sokol \mcite{KvamSokol06} for college basketball, where the 
best $p$ transition probabilities include information about point spread and home-court advantage.  Finally, we might wonder whether some optimization would connect this class of ``direct
methods'' \mcite{Keener93} to the more statistically-sophisticated
maximum likelihood methods originally used for paired comparisons in the
pioneering works of Zermelo \mcite{Zermelo26} or Bradley and Terry 
\mcite{BradleyTerry52}. Rather than continuing these lines of inquiry here, we instead focus our remaining attention on trying to understand the connections between our biased random walkers and the properties of the underlying networks.

\section{NETWORK STRUCTURE.}    
\label{sec:network}

The voting automatons randomly walk on a graph consisting of the
Division I-A football teams (vertices) connected by the games
played between the teams (edges).  We note that this is not the only ranking system motivated by direct interest in the underlying network; indeed, while the random walkers here provide rankings through their dynamics on the network, Park and Newman
\mcite{ParkNewman05} have developed a ranking system determined completely by the directed graph defined by wins.  Because the random walkers propagate in a strongly heterogeneous network topology, it is important to try to understand how that topology affects the walkers' behavior and the resulting rankings.

Each NCAA Division I-A football season consists of
650--750 games between about 115 teams (119 in 2005).  In every season since
1990, this graph becomes a single connected component by the third or
fourth week of the season (without counting faux connections via the
single ``non-I-A'' node that we include for simplicity).
The relative quickness in achieving a single connected
component (after only about two hundred games) results in part because schools
typically play many of their nonconference games at the beginning of
a season.  The degree of each vertex---that is, the number of games played by each
team---varies in a narrow range.  Each vertex has between ten and
thirteen edges prior to
the bowl games and between ten and fourteen connections after the bowl
games. The diameter of the graph, determined by counting the number of edges along the longest geodesic path, is 4 in every post-bowl graph since 1970. We also considered
the local clustering coefficient $C_i$ for the $i$th team in the network 
(see \mcite{WattsStrogatz98}, \mcite{NewmanSIREV}), given by
\begin{equation}
  C_i = \frac{\mbox{number of triangles connected to vertex } i}
  {\mbox{number of triples centered on vertex } i}\,.
\end{equation}
Unsurprisingly, clustering coefficients provide one means of
identifying the strong heterogeneity of the network.  Each conference
typically has a different average local clustering coefficient, which
also varies from the coefficients computed for independent teams such
as Notre Dame and Navy.

Several other network properties can also be
calculated (including average path lengths and various notions of centrality and connectedness; see, for example, \mcite{NewmanSIREV}), but such computations do not necessarily help explain the
random-walker statistics. An important exception is the graph's
{``community structure''} \mcite{GirvanNewman02}, which indicates the
hierarchies present in the network and is useful for understanding the
nature of the conference scheduling and the resulting effect on the
random-walker statistics.  We computed community structure using the
notion of ``edge betweenness,'' defined as the number of geodesics
that traverse each edge, using the algorithm given by
\mcite{GirvanNewman02}.  Briefly, the edge with the highest betweenness
is removed from the graph, and the betweenness is recalculated for this modified
graph to determine the next edge to remove; the process is then repeated until no edges remain.  The removal of some of these edges breaks a connected component of the graph into two parts,
grouping the network into a hierarchy of communities as the algorithm
is iterated. Girvan and Newman \mcite{GirvanNewman02} demonstrated that
their algorithm closely reproduced the predefined conference structure
of the 2000 football schedule, and we found similar results for other
years.

\begin{figure}[ht]
  \centerline{
    \includegraphics[width=\textwidth]{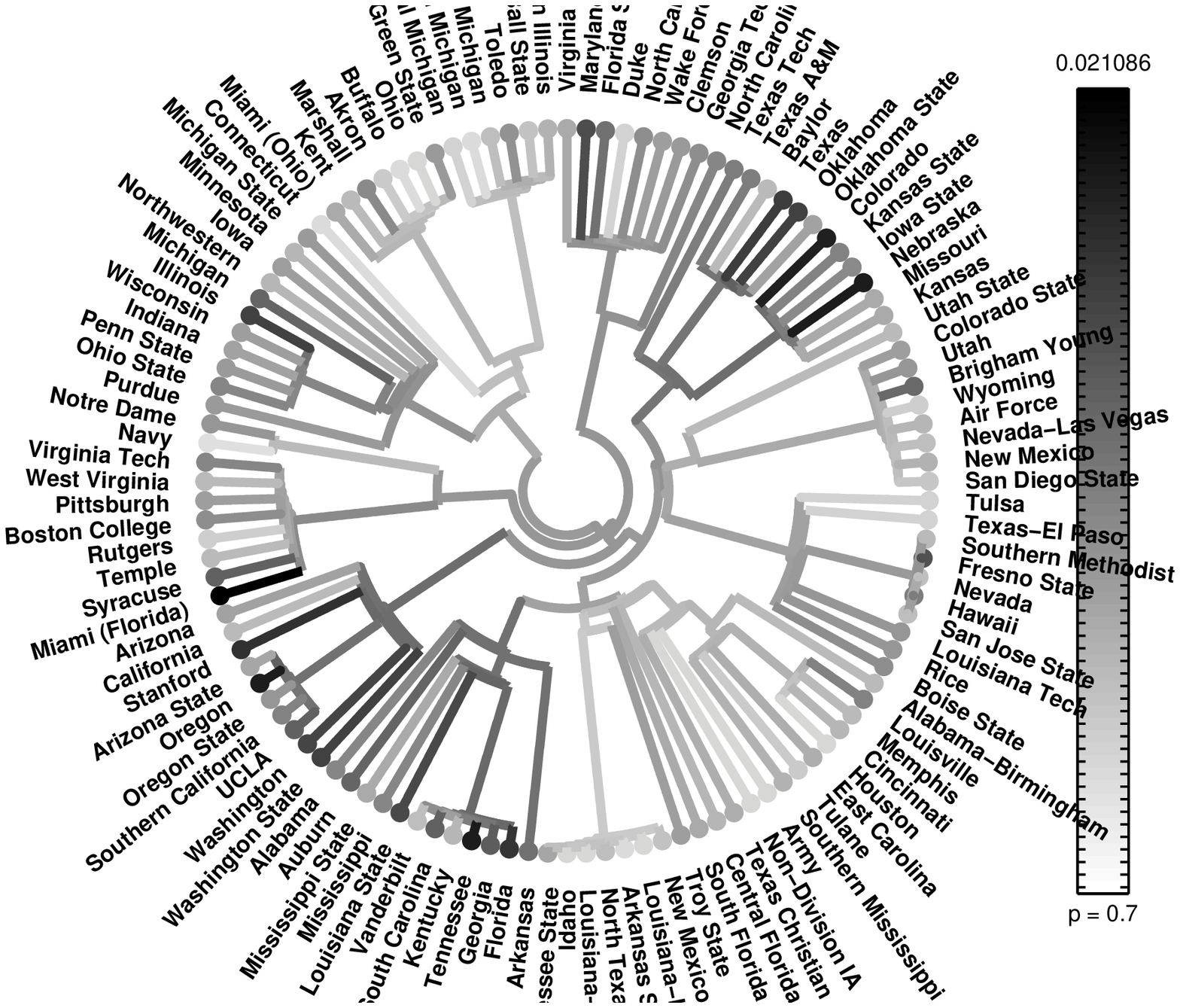}
  }
  \caption{Graphical depiction of the community structure of the 
    2001 pre-bowl network at $p = 0.7$, with gray-scale coding of 
    the average votes per team for each community.}
  \label{fig:community}
\end{figure}

Using the 2001 season as an example, Figure \ref{fig:community}
portrays the college football community structure. The conferences and
their subdivisions are reasonably reconstructed based on their relative
strengths of community, and the tree (``dendrogram") in the figure indicates the
relative closeness of different conferences and of the independent
teams. Note, for example, the close connections between Notre Dame,
Navy, and the Big East (from Virginia Tech counterclockwise to Miami (Florida)).

This community structure is intimately linked to the dynamics of the
random-walking voters, as the specific pairings of interconference
games and the outcomes of those games strongly influence the
flow of voters into and out of the given conferences and more
general structures such as the divisions inside large conferences and
larger hierarchical groups of conferences. The 2001 pre-bowl community
structure in Figure \ref{fig:community} is gray-scaled
according to the average number of voters per team
at each level of the hierarchy (for $p=0.7$). Such a plot indicates the
relatively high vote counts given to the SEC, Pacific-10 (Pac-10), and
Big 12; it also shows the significantly smaller average vote counts for Big East teams, despite Miami's first-place standing.

\begin{figure}[ht]
  {\psfragscanon\footnotesize
    \psfrag{Propagated Difference Squared}[b][b]{Effect of reversal}
    \psfrag{Order of edge removal}[t][t]{Order of edge removal}
    \centerline{
      \includegraphics[width=0.8\textwidth]{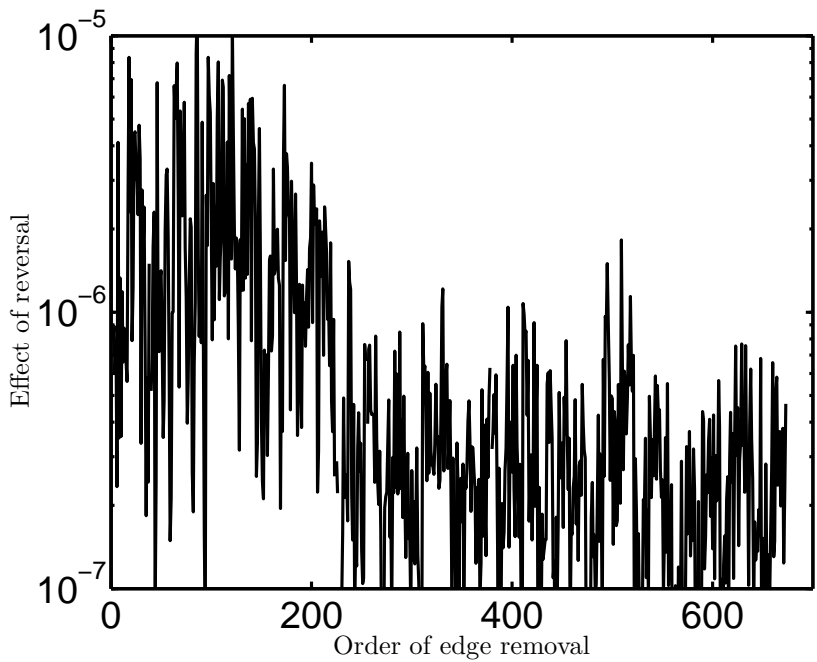}
    }
  }
  \caption{The effect of reversing
    the outcome of a single game played between teams $j$ and $k$,
    quantified by $\sum_{i\neq j,k} |\bar{v}_i|^2$ (for $p=0.7$) and 
    plotted versus the order of that game according to edge 
    removal in the community structure determination.}
  \label{fig:ags}
\end{figure}

We further quantify the importance of the relationships between
conferences by directly measuring the effect of reversing the outcomes
of individual games. Keeping the games in the order in which their
respective edges were removed in developing the community structure,
we measure the difference between the original voter populations and
the new populations calculated with the win-loss outcome of that
single game reversed.  The dominant effect of such reversals is to
change the rankings of the two teams involved in that game, so we
calculate the change to the global distribution of votes with the
quantity $\sum_{i\neq j,k}|\bar{v}_i|^2$, where the edge removed
corresponds to a game between teams $j$ and $k$. Plotting this
quantity (see Figure \ref{fig:ags}) versus the order of the edge removed
reveals a sharp transition in average magnitude between the first
approximately two hundred edges and those that follow. This corresponds
roughly to the number of edges removed in the community structure
determination at which the hierarchy breaks up into the different
conferences; these first couple hundred edges removed are predominantly interconference games,
whereas those that follow are intraconference games

\section{DISCUSSION.}    
\label{sec:summary}

We have developed a simply-defined ranking algorithm in
which random walkers adjust their ``votes'' for the best team on the 
network of teams connected by their head-to-head games, with a single explicit
parameter describing the preference towards selecting the winner of a given game. We investigated the asymptotic behavior of this algorithm in the two extremes of the probability
parameter, paying special attention to round-robin tournaments, and explored the results of the 2001 and 2002 NCAA Division I-A college football seasons in detail.  More recent seasons are discussed on our website \mcite{ourweb}.  Finally, we have connected the rankings to the underlying network of games played by Division I-A teams, quantitatively demonstrating the importance of interconference game outcomes and the relation to the community structure of the graph.

Of course, many generalizations of the random-walker ranking algorithm described here are possible, 
ranging from trivial redefinitions of the rate matrix $\mathbf{D}$ to
fundamental changes in the mathematical tools required to investigate
the random-walking voters. 
The simplest generalizations are those that modify the rate matrix
$\mathbf{D}$ without changing the independence of the random walkers
themselves. For instance, one can readily incorporate margin of
victory, home-field advantage, and game date into the definition of
$\mathbf{D}$ by replacing the constant probability $p$ of voting for
the winner with a function that includes these components (see, for example, \mcite{KvamSokol06}). The resulting
transition rates of walking each direction on a single edge are still
defined by the outcome of the game represented by that edge, and the
associated linear algebra problem determines the probability
distribution of each state.  It would be natural to make the
probability of going towards the winner along a given edge higher for
a larger margin of victory or for a game won on the road. However, incorporating any such qualitative assertions into quantitative variations in the transition rates
requires adding more
parameters to the ranking algorithm and determining reasonable values of those parameters, in stark contrast to the minimalist philosophy espoused here.

Alternatively, each random walker can be assigned two votes instead
of one.  This change is particularly sensible if the point is to
select the two teams to play head-to-head in the National Championship
Game. We considered such rankings generated from random walkers, each of whom holds two equal votes (as opposed to a \#1 vote and a separate \#2
vote), using the same probability parameter $p$, subject to the
additional constraint that a given voter must cast the votes for
two different teams. This two-vote constrained random walk is most
easily understood in terms of independent random walks on the
(significantly larger!) network in which each vertex represents a
possible pair of votes and the edges between vertices include games
played between teams representing one of the two votes, with the other
vote held fixed. This again immediately reduces to a linear algebra
problem for the expected percentage of votes garnered by each team,
although it has much higher rank: $T(T-1)/2$ for $T$ teams. Clearly,
further increases in the number of votes given to each random walker
would quickly make the state space so large that solving the exact
linear algebra problem would no longer be computationally feasible.

As examples, we considered this two-vote generalization for pre-bowl rankings for both 2001 and 2002.  Our results were unremarkably similar to our single-vote rankings. In particular, in comparison with the single-vote 2001 case, Miami can (of course) obtain only
half the votes in the limit as $p\to 1$ because of the constraint that
each walker casts the two votes for two different teams, leaving fifty percent
of the votes available to select a \#2 team.  That fifty percent becomes widely divided among the available candidates, with Tennessee and Oregon getting the largest shares at roughly five percent each. Meanwhile, this major change in the expected populations causes only relatively small changes in the rank orderings, except at extreme values of $p$, where Tennessee and the rest of the SEC do not fare quite as well here near $p=1$ as they do in the single-vote scheme (though Tennessee still
maintains the \#2 spot at most values of $p$).  At $p=0.75$, the single-vote and two-vote systems agree on the same top sixty teams in the 2001 pre-bowl rankings, save for swaps in the orderings of \#19-20 and \#48-49; at $p=0.9$, the top twenty-five are identical; at $p=0.95$, however, there are switches in the orderings of \#3-4, \#5-6, and \#9-10 (SEC teams Florida, LSU, and Georgia all do better in the single-vote algorithm), while the rest of the top twenty-five remains identical. Not unexpectedly,
the 2002 two-vote rankings are very similar to the single-vote values, with Miami and Ohio State splitting the votes in the limit as $p\to 1$.  Even at $p=0.95$, the top eighteen teams are identical except for a reversal in the ordering of \#6-7.

One can also consider generalizations that destroy the independence of
individual random walkers.  For example, voter decisions could be
influenced by the number of other walkers voting for each team in a
head-to-head game. Whether they are inclined to follow the crowd or to
try to be nonconformist, such dependence between the random walkers
makes the calculation of their aggregate behavior nonlinear and
removes most of our knowledge about the probability distributions of
votes per team. Another interesting generalization would be to weigh more strongly the effects of upsets by increasing the flow of votes towards a
lower-ranked team that beats a team with more votes.  Of course, this
flow increase might reverse the ordering of the two teams, thereby
removing the upset character and reducing the flow towards the winning
team of that game, so there may not even be a statistically-steady
ordering of the two teams. A similar complication occurs if one breaks
the independence of the walkers by adding a bonus for beating a team
ranked in (for example) the top ten that reduces the probability of
voting for the loser. This reduced flow towards the losing team could then
knock them out of the top ten, thereby causing the bonus to disappear,
and allowing the team to rise back into the top ten, and so on. Obviously, the
study of any of these interacting random walkers is significantly more
difficult than the independent walkers considered in this article.

As a last consideration, we return to our discussion in section 
\ref{sec:roundrobin} of crossing probabilities $p_c$ for round-robin competition and the observation that $p_c$ can be very close to $1/2$ when a team with a worse record won its games against quality opponents, such as the specific mixed ordering we considered. The random walkers then reward a team more for its high-quality wins than they penalize it for its low-quality losses. This issue can be easily corrected by introducing a new ranking according to the difference of expected vote populations obtained with parameter $p$ (``first-place votes'' rewarding high-quality wins) and those obtained with $1-p$ (``last-place votes'' penalizing low-quality losses). For the specific mixed ordering considered in section \ref{sec:roundrobin} this ``first-last'' generalization gives $p_c \gtrsim .77$ for large $T$, while Monte Carlo searches preliminarily suggest that 
$p_c$ may indeed be bounded away from $1/2$ for these first-last random walker rankings for round-robin tournaments. While this improvement is mathematically more satisfying than the situation for the original ``first-only'' rankings, the first-only and first-last rankings nevertheless generally agree on the top teams for real football seasons.  For instance, in the recently-concluded 2005 season these two pre-bowl rankings at $p=0.75$ agree on the ordering of the top twelve teams except for a \#10-11 swap, with the revised (``first-last") system giving a slightly smaller number of ranking violations than the first-only system (complete rankings are available from 
\mcite{ourweb}). 

After eight seasons, the Bowl Championship Series remains as controversial as ever. Even when the system yields an uncontroversial National Championship Game because two teams clearly separate themselves from the field (as in 2002 and 2005), the BCS is still unable to escape controversy. The treatment of the so-called mid-major (or ``non-BCS'') conferences remains an important issue 
(see, for example, \mcite{Callaghan04}), leading to a December 2005 Congressional hearing in a House Energy and Commerce subcommittee \mcite{Jenkins05}. We remain committed to the proposition that the use of algorithmic rankings for determining the college football postseason will only become widely accepted when those rankings have been reasonably explained to the public. In that context, the random walker rankings (and their first-last generalization) provide reasonable ways to rank teams algorithmically with methods that can be easily explained and broadly understood.

\bigskip\noindent
{\bf ACKNOWLEDGMENTS.}
We are grateful to both Kenneth Massey and Mark Newman for several discussions over the
course of this project, to Michael Abraham for developing some of the
codes used in our analysis, and to the anonymous referees for their
helpful criticisms. This work was supported in part by an NSF VIGRE grant (DMS-0135290) to the Georgia Institute of Technology, where this work began.

\providecommand{\bysame}{\leavevmode\hbox to3em{\hrulefill}\thinspace}
\providecommand{\MR}{\relax\ifhmode\unskip\space\fi MR }
\providecommand{\MRhref}[2]{%
  \href{http://www.ams.org/mathscinet-getitem?mr=#1}{#2}
}
\providecommand{\href}[2]{#2}


\begin{thebibliography}{10}

\bibitem{BCS}
{Bowl Championship Series}, http://www.bcsfootball.org, updated 2006.

\bibitem{BradleyTerry52}
R.~A. Bradley and M.~E. Terry, Rank analysis of incomplete block designs.
  {I. The} method of paired comparisons, \emph{Biometrika} \textbf{39} (1952)
  324--345.

\bibitem{BrinPage99}
S.~Brin, L.~Page, R.~Motwami, and T.~Winograd, The PageRank citation
  ranking: Bringing order to the web, Tech. Report 1999-0120, Computer
  Science, Stanford University, Stanford, CA, 1999.

\bibitem{Callaghan04}
T.~Callaghan, P.~J. Mucha, and M.~A. Porter, The Bowl Championship
  Series: A mathematical review, \emph{Notices Amer. Math. Soc.}  \textbf{51}
  (2004) 887--893.

\bibitem{Coleman05}
J.~Coleman, Minimizing game score violations in college football
  rankings, \emph{Interfaces} \textbf{35} (2005) 483--496.

\bibitem{ColleyWeb}
W.~N. Colley, ColleyMatrix: Colley's bias free matrix rankings,
  updated 2006, http://www.colleyrankings.com/.

\bibitem{ConnorGrant00}
G.~R. Connor and C.~P. Grant, An extension of Zermelo's model for
  ranking by paired comparisons, \emph{Eur. J. App. Math.} \textbf{11} (2000)
  225--247.

\bibitem{GirvanNewman02}
M.~Girvan and M.~E.~J. Newman, Community structure in social and
  biological networks, \emph{Proc. Nat. Acad. Sci} \textbf{99} (2002) 7821--7826.

\bibitem{Howell}
J.~Howell, James Howell's college football scores, downloaded 2003,
  \hfill
  http://homepages.cae.wisc.edu/{\small$\sim$}dwilson/rsfc/history/howell/.

\bibitem{Jenkins05}
S.~Jenkins, BCS schemers are looking out for no.\ 1, \emph{Washington Post}
  (10 December, 2005), E01.

\bibitem{Keener93}
J.~P. Keener, The Perron-Frobenius theorem and the ranking of
  football teams, \emph{SIAM Rev.}  \textbf{35} (1993) 80--93.

\bibitem{KvamSokol06}
P. Kvam and J.S. Sokol, A Logistic regression/Markov chain model for NCAA basketball,  \emph{Naval Research Logistics}  (2006)
(to appear); available at
http://www2.isye.gatech.edu/{\small$\sim$}jsokol/ncaa.pdf. 

\bibitem{LangvilleMeyer05}
A.~N. Langville and C.~D. Meyer, A survey of eigenvector methods for web
  information retrieval, \emph{SIAM Review} \textbf{47} (2005) 135--161.

\bibitem{Martinich02}
J.~Martinich, College football rankings: Do the computers know best?,
  \emph{Interfaces} \textbf{32} (2002) 85--94.

\bibitem{MasseyComparisons}
K.~Massey, College football ranking comparison, updated 2006,
  http://www.masseyratings.com/cf/compare.htm.

\bibitem{ourweb}
P.~J. Mucha, T.~Callaghan, and M.~A. Porter, Random walker rankings
  for NCAA Division I-A football, updated 2006,
  http://www.amath.unc.edu/Faculty/mucha/BCS/.

\bibitem{NewmanSIREV}
M.~E.~J. Newman, The structure and function of complex networks, 
\emph{SIAM  Rev.} \textbf{45} (2003) 167--256.

\bibitem{Park05}
J.~Park, On minimum violations ranking in paired comparisons,
(2005), http://www.arxiv.org/abs/physics/0510242.

\bibitem{ParkNewman05}
J.~Park and M.~E.~J. Newman, A network-based ranking system for US
  college football, \emph{J. Stat. Mech.} (2005) P10014.

\bibitem{Stefani97}
R.~T. Stefani, Survey of the major world sports rating systems, 
\emph{J. App.  Stats.} \textbf{24} (1997) 635--646.

\bibitem{WattsStrogatz98}
D.~J. Watts and S.~H. Strogatz, Collective dynamics of ``small-world''
  networks, \emph{Nature} \textbf{393} (1998) 440--442.

\bibitem{dwilson}
D.~L. Wilson,
http://homepages.cae.wisc.edu/\small{$\sim$}dwilson/rsfc/rate/,
  updated 2006.

\bibitem{Wolfe}
P.~R. Wolfe, http://prwolfe.bol.ucla.edu/cfootball/, downloaded
  2003--2005.

\bibitem{Zermelo26}
E.~Zermelo, Die Berechnung der Turnier-Ergebnisse als ein
  Maximum-problem der Wahrscheinlichkeitsrechnung, 
  \emph{Math. Z.} \textbf{29} (1926) 436--460.

\end{thebibliography}

\bigskip

\noindent{\bf THOMAS CALLAGHAN} is a graduate student in the Institute for Computational and Mathematical Engineering at Stanford University. He is a 2005 graduate of the Georgia Institute of Technology, where he started this work as an REU project at a time when all three authors were in the School of Mathematics there. He won his girlfriend's heart after she used this ranking system to win a college football bowl game pool among her coworkers.  A fan of many sports, he is an avid ultimate frisbee player, when he is not too busy with research or autographing the November 2004 \emph{ESPN: The Magazine} in which this work was featured. He is the only coauthor here with a degree from a Division I-A football school.\\
\emph{Institute for Computational and Mathematical Engineering,\\
Stanford University, Stanford, CA 94305-4042 USA\\
tscallag@stanford.edu}

\bigskip

\noindent{\bf PETER MUCHA} is an assistant professor in the Department of Mathematics and the Institute for Advanced Materials at the University of North Carolina at Chapel Hill. Previously, he was an assistant professor in the School of Mathematics at Georgia Tech. He received a Ph.D. from Princeton University in 1998. While he greatly enjoys college football, as a graduate of Cornell University he is naturally a much bigger fan of college hockey, which manages to seed its annual sixteen-team postseason tournament by a committee decision that is almost wholly dictated algorithmically (see, for example, http://www.uscho.com).\\
\emph{Department of Mathematics \& Institute for Advanced 
Materials,\\
University of North Carolina, Chapel Hill, NC 27599-3250 USA\\
mucha@unc.edu}

\bigskip

\noindent{\bf MASON PORTER} is a postdoctoral scholar in the Center for the Physics of Information and the Department of Physics at California Institute for Technology. Prior to his 2005 return to his beloved alma mater (B.S., Applied Mathematics, 1998), which also brought him closer to his equally beloved Los Angeles Dodgers, he received a Ph.D. from the Center for Applied Mathematics at Cornell University in 2002 and was a VIGRE Visiting Assistant Professor in the School of Mathematics at Georgia Tech.  He would like to point out that Caltech's football team has gone undefeated since 1993 (when it was disbanded).  Most of its existing teams haven't won any games in a while either.\\
\emph{Department of Physics \& Center for the Physics of 
Information,\\
California Institute of Technology, Pasadena, CA  91125-3600 USA\\
mason@its.caltech.edu}

\end{document}